\documentclass[aps,prd,twocolumn,superscriptaddress,nofootinbib,showpacs]{revtex4-1}

\usepackage{amsmath,bbm,latexsym,amssymb,multirow,graphicx,float}

\def\be{\begin{equation}}
\def\ee{\end{equation}}
\def\ba{\begin{eqnarray}}
\def\ea{\end{eqnarray}}

\begin{document}

\title{Models with three Higgs doublets in the triplet representations of $A_4$ or $S_4$}

\author{R. Gonz\'{a}lez Felipe}\thanks{E-mail: ricardo.felipe@ist.utl.pt}
\affiliation{Instituto Superior de Engenharia de Lisboa - ISEL,
	1959-007 Lisboa, Portugal}
\affiliation{Centro de F\'{\i}sica Te\'{o}rica de Part\'{\i}culas (CFTP),
    Instituto Superior T\'{e}cnico, Universidade T\'{e}cnica de Lisboa,
    1049-001 Lisboa, Portugal}
\author{H.~Ser\^{o}dio}\thanks{E-mail: hugo.serodio@ific.uv.es}
\affiliation{Departament de F\'{\i}sica Te\`{o}rica and IFIC,
		Universitat de Val\`{e}ncia-CSIC,
		E-46100, Burjassot, Spain}
\author{Jo\~{a}o P.~Silva}\thanks{E-mail: jpsilva@cftp.ist.utl.pt}
\affiliation{Instituto Superior de Engenharia de Lisboa - ISEL,
	1959-007 Lisboa, Portugal}
\affiliation{Centro de F\'{\i}sica Te\'{o}rica de Part\'{\i}culas (CFTP),
    Instituto Superior T\'{e}cnico, Universidade T\'{e}cnica de Lisboa,
    1049-001 Lisboa, Portugal}

\date{\today}

\begin{abstract}
We consider the quark sector of theories containing three scalar $SU(2)_L$
doublets in the triplet representation of $A_4$ (or of $S_4$) and three
generations of quarks in arbitrary $A_4$ (or $S_4$) representations. We show
that, for all possible choices of quark field representations and for all
possible alignments of the Higgs vacuum expectation values that can constitute
global minima of the scalar potential, it is not possible to obtain
simultaneously non-vanishing quark masses and a non-vanishing CP-violating
phase in the CKM quark mixing matrix. As a result, in this minimal form,
models with three scalar fields in the triplet representation of $A_4$ or of
$S_4$ cannot be extended to the quark sector in a way consistent with experiment.
\end{abstract}


\maketitle

\section{\label{sec:intro}Introduction}

There is a long history of articles considering discrete symmetries in the study
of the leptonic sector (see for instance the recent
reviews~\cite{Altarelli:2010gt,Ishimori:2010au,Grimus:2011fk,King:2013eh}
and references therein), including many models predicting tri-bimaximal
leptonic mixing~\cite{Harrison:2002er}, now disfavored by the measurement
of a large mixing angle $\theta_{13}$~\cite{Abe:2011sj,Abe:2011fz,An:2012eh,Ahn:2012nd}.
In the quark sector, models based on the $A_4$ symmetry as a
possible family symmetry were first introduced in
Refs.~\cite{Wyler:1979fe,Branco:1980ax}.
After the impact of the symmetry on the Yukawa matrices is known,
some structure for the vacuum expectation values (vev) has to be
assumed before moving on to the mass matrices and respective
phenomenological predictions. Occasionally, this has been
performed without a full study of the scalar sector and without
ensuring properly whether the assumed vacuum structure indeed
corresponds to the global minimum. This may occur in part because
finding local minima is easy (one just has to show that the gradient
of the potential vanishes), while ensuring that there is no other,
lower-lying, minimum is often rather difficult.
Recently, Degee, Ivanov, and Keus~\cite{ivanov_min}
have introduced a geometrical procedure to minimize highly
symmetric scalar potentials, and solved the problem for a
three Higgs doublet model (3HDM) potential with an $A_4$
or an $S_4$ symmetry.
Although it is not explicitly stated,
Ref.~\cite{ivanov_min} refers to a set of three Higgs fields
in a triplet representation of the
group\footnote{To be precise, the three scalar fields must
be in a faithful representation of the group~\cite{ivanov-private}.}.
This is a crucial point since,
if one were to place each of the three Higgs fields
in a singlet representation, then one would end up
with the most general 3HDM potential.
It is found that the possible vev alignments
for the $A_4$ symmetric potential \cite{merlo} which may
correspond to a global minimum are~\cite{ivanov_min}
\begin{eqnarray}
&&v\, (1,0,0),\nonumber\\
&&v\, (1,1,1),\nonumber\\
&&v\, (\pm 1, \eta, \eta^*)\ \textrm{with }\ \eta = e^{i \pi/3},\nonumber\\
&&v\, (1, e^{i \alpha}, 0)\ \textrm{with any}\ \alpha.
\label{vev:A4}
\end{eqnarray}
Similarly, the possible vev alignments corresponding
to global minima in the $S_4$ symmetric potential are~\cite{ivanov_min}
\begin{eqnarray}
&&v\, (1,0,0),\nonumber\\
&&v\, (1,1,1),\nonumber\\
&&v\, (\pm 1, \eta, \eta^*)\ \textrm{with }\ \eta = e^{i \pi/3},\nonumber\\
&&v\, (1, i, 0).
\label{vev:S4}
\end{eqnarray}
In each case, a vev corresponding to some permutation of
the fields is also a possible global minimum.
Any other solution of the stationarity conditions may be a saddle point,
a local maximum, or even a local minimum, but never the global minimum.

Besides a correct identification of global minima,
one must also consider whether the specific discrete symmetry
under study can be extended to the whole Lagrangian of the theory,
in a way consistent with known data.
In particular, in the quark sector there should be no
massless quarks, no diagonal blocks in the CKM matrix,
and/or no vanishing CP-violating phase.
As shown by Ferreira and Silva~\cite{FS3},
these constraints place stringent limits on the
type of mass matrices obtainable from Abelian symmetries in the 2HDM.

In this article, we consider models with three Higgs doublets
$\Phi_i$ in a triplet representation of $A_4$ (Sec.~\ref{sec:A4}),
or in a triplet representation of $S_4$ (Sec.~\ref{sec:S4}).
This ensures that the only possible global vev structures are
those in Eqs.~\eqref{vev:A4} and \eqref{vev:S4}, respectively.
The models contain only three generations of left-handed quark
doublets $Q_L$, right-handed up-type quark singlets $u_R$,
and right-handed down-type quark singlets $d_R$.
Our conclusions are briefly summarized in Sec.~\ref{sec:conclusions}.

\section{\label{sec:A4}The $A_4$ case}

$A_4$ is the group of the even permutations of four objects
and it has 12 elements divided into four irreducible representations,
namely, three singlets
$\ \mathbf{1}, \ \mathbf{1^{\prime}}, \ \mathbf{1^{\prime \prime}}$
and one triplet $\ \mathbf{3}$.
The multiplication rules are
\begin{eqnarray} \label{A4rules}
\mathbf{1} \otimes \textrm{any}
&=&
\textrm{any},
\nonumber\\
\mathbf{1^\prime} \otimes \mathbf{1^\prime}
&=&
\mathbf{1^{\prime \prime}},
\nonumber\\
\mathbf{1^\prime} \otimes \mathbf{1^{\prime \prime}}
&=&
\mathbf{1},
\nonumber\\
\mathbf{1^\prime} \otimes \mathbf{3}
&=&
\mathbf{3},
\\
\mathbf{1^{\prime \prime}} \otimes \mathbf{1^{\prime \prime}}
&=&
\mathbf{1^\prime},
\nonumber\\
\mathbf{1^{\prime \prime}} \otimes \mathbf{3}
&=&
\mathbf{3},
\nonumber\\
\mathbf{3} \otimes \mathbf{3}
&=&
\mathbf{1} \oplus \mathbf{1^\prime} \oplus \mathbf{1^{\prime \prime}}
\oplus \mathbf{3_s}
\oplus \mathbf{3_a}.\nonumber
\end{eqnarray}

We recall that, for the corresponding entry of the Yukawa coupling matrix to
be non-vanishing, the Yukawa Lagrangian must be in the invariant singlet representation
$\mathbf{1}$ . Since the three Higgs doublets are in the representation $\mathbf{3}$,
we see from Eqs.~\eqref{A4rules}, that the product of left-handed and right-handed
fermions must also be in a triplet representation. This means that at least one of
the fermion fields in each charge sector must be in a triplet representation.
The possibilities for the representations of the left-handed quark fields and
for the up and down right-handed quarks are listed in Table~\ref{tabA4:rep}.
\begin{table}[h!]
\begin{center}
\begin{tabular}{|ccccccc|}
\hline
& $Q_L$ & & $u_R$ & & $d_R$ & \\
\hline
& $\mathbf{3}$  & &
$\mathbf{3}$  & &
$\mathbf{3}$ & \\
& $\mathbf{3}$  & &
$\mathbf{3}$  & &
three singlets & \\
& $\mathbf{3}$  & &
three singlets  & &
$\mathbf{3}$ & \\
& $\mathbf{3}$  & &
three singlets & &
three singlets & \\
& three singlets  & &
$\mathbf{3}$  & &
$\mathbf{3}$ & \\
\hline
\end{tabular}
\end{center}
\caption{Possible representations of the left-handed quark doublets ($Q_L$),
the right-handed up quark singlets ($u_R$), and the right-handed down quark
singlets ($d_R$), when the three Higgs doublets are in a triplet representation
$\mathbf{3}$.
\label{tabA4:rep}}
\end{table}

Since permutations of the three fields in each sector do not lead to new
structures for the Yukawa matrices, the notation ``three singlets'' stands
for the following independent possibilities for the fields in each of the
three generations:
\begin{eqnarray}
(\mathbf{1}, \mathbf{1}, \mathbf{1}),
& \quad &
(\mathbf{1}, \mathbf{1^\prime}, \mathbf{1^{\prime \prime}}),
\nonumber\\
(\mathbf{1}, \mathbf{1}, \mathbf{1^\prime}),
& \quad &
(\mathbf{1^\prime}, \mathbf{1^\prime}, \mathbf{1^\prime}),
\nonumber\\
(\mathbf{1}, \mathbf{1^\prime}, \mathbf{1^\prime}),
& \quad &
(\mathbf{1^\prime}, \mathbf{1^\prime}, \mathbf{1^{\prime\prime}}),
\nonumber\\
(\mathbf{1}, \mathbf{1}, \mathbf{1^{\prime \prime}}),
& \quad &
(\mathbf{1^\prime}, \mathbf{1^{\prime\prime}}, \mathbf{1^{\prime\prime}}),
\nonumber\\
(\mathbf{1}, \mathbf{1^{\prime \prime}}, \mathbf{1^{\prime \prime}}),
& \quad &
(\mathbf{1^{\prime\prime}}, \mathbf{1^{\prime\prime}}, \mathbf{1^{\prime\prime}}).
\end{eqnarray}

In order to use the vevs given in Eq.~\eqref{vev:A4}, one must be sure to use
a representation of the group that is consistent with the basis in which those
vevs were obtained in Ref.~\cite{ivanov_min}. Indeed, if one starts from Higgs
fields with the vevs of Eq.~\eqref{vev:A4}, and one changes the scalar fields
by a unitary transformation $U$, i.e.
\begin{equation}
\left(
\begin{array}{c}
\Phi_1\\
\Phi_2\\
\Phi_3
\end{array}
\right)
\rightarrow
U
\left(
\begin{array}{c}
\Phi_1\\
\Phi_2\\
\Phi_3
\end{array}
\right),
\end{equation}
then the vevs also transform as
\begin{equation}
\left(
\begin{array}{c}
\langle\Phi_1\rangle\\
\langle\Phi_2\rangle\\
\langle\Phi_3\rangle
\end{array}
\right)
\rightarrow
U
\left(
\begin{array}{c}
\langle\Phi_1\rangle\\
\langle\Phi_2\rangle\\
\langle\Phi_3\rangle
\end{array}
\right),
\end{equation}
and, in general, will no longer have the form in Eq.~\eqref{vev:A4}. A suitable
basis for the triplet representation of $A_4$ is given by
\begin{equation}
S =
\left(
\begin{array}{ccc}
1 & 0 & 0\\
0 & -1 & 0\\
0 & 0 & -1
\end{array}
\right),
\hspace{10mm}
T =
\left(
\begin{array}{ccc}
0 & 1 & 0\\
0 & 0 & 1\\
1 & 0 & 0
\end{array}
\right).
\label{S&T}
\end{equation}
In the notation of Sec.~6.4 of Ref.~\cite{Ivanov_Vdovin_2}, $a_1 = S$, $b = T$,
and $a_2 = T^{-1} S T$ is redundant. These matrices satisfy $S^2=T^3=(ST)^3 = 1$,
showing that they indeed generate the group $A_4$. Equations~\eqref{S&T} also
coincide with the basis used in Ref.~\cite{lepto}.

One way to confirm that we are indeed using a basis consistent with
Ref.~\cite{ivanov_min} is to check that imposing $S$ and $T$ on the 3HDM
potential, we recover
\begin{eqnarray}
V =
&-& \frac{M_0}{\sqrt{3}}
\left( |\Phi_1|^2 + |\Phi_2|^2 + |\Phi_3|^2 \right)\nonumber\\
&+&
\frac{\Lambda_0}{3}
\left( |\Phi_1|^2 + |\Phi_2|^2 + |\Phi_3|^2 \right)^2
\nonumber\\
&+&
\frac{\Lambda_3}{3}
\left[
|\Phi_1|^4 + |\Phi_2|^4 + |\Phi_3|^4\right.\nonumber\\
& &\left. -|\Phi_1|^2 |\Phi_2|^2 - |\Phi_2|^2 |\Phi_3|^2 - |\Phi_3|^2 |\Phi_1|^2
\right]\nonumber\\
&+&
\Lambda_1
\left[
(\textrm{Re} \Phi_1^\dagger \Phi_2)^2
+ (\textrm{Re} \Phi_2^\dagger \Phi_3)^2
+ (\textrm{Re} \Phi_3^\dagger \Phi_1)^2
\right]
\nonumber\\
&+&
\Lambda_2
\left[
(\textrm{Im} \Phi_1^\dagger \Phi_2)^2
+ (\textrm{Im} \Phi_2^\dagger \Phi_3)^2
+ (\textrm{Im} \Phi_3^\dagger \Phi_1)^2
\right]
\nonumber\\
&+&
\Lambda_4
\left[
(\textrm{Re} \Phi_1^\dagger \Phi_2)(\textrm{Im} \Phi_1^\dagger \Phi_2)
+ (\textrm{Re} \Phi_2^\dagger \Phi_3)(\textrm{Im} \Phi_2^\dagger \Phi_3)\right.\nonumber\\
& & \left.+ (\textrm{Re} \Phi_3^\dagger \Phi_1)(\textrm{Im} \Phi_3^\dagger \Phi_1)
\right],
\label{masterV}
\end{eqnarray}
as in Eq.~(9) of Ref.~\cite{ivanov_min}~\footnote{Equation~(9) of
Ref.~\cite{ivanov_min} coincides with the sum of Eqs.~(38) and (39) of
Ref.~\cite{Ivanov_Vdovin_2}, with the substitutions
$\Lambda_0 = 3 \lambda + \lambda^\prime,\quad \Lambda_1 = \lambda^{\prime \prime}
+ 2 \textrm{Re}(\tilde{\lambda}),\quad \Lambda_2 = \lambda^{\prime \prime}
- 2 \textrm{Re}(\tilde{\lambda}),
\quad \Lambda_3 = - \lambda^\prime, \quad \Lambda_4
= - 4  \textrm{Im}(\tilde{\lambda})$.}.

In $A_4$, with the basis of Eq.~\eqref{S&T}, the product of two triplets,
$a=(a_1, a_2, a_3)$ and $b=(b_1, b_2, b_3)$, gives~\cite{lepto,Altarelli:2010gt}
\begin{eqnarray}
&&(a \otimes b)_{\mathbf{1}} =
a_1 b_1 + a_2 b_2 + a_3 b_3,
\label{33:1}
\nonumber\\
&&(a \otimes b)_{\mathbf{1^\prime}} =
a_1 b_1 + \omega^2 a_2 b_2 + \omega a_3 b_3,
\label{33:1prime}
\nonumber\\
&&(a \otimes b)_{\mathbf{1^{\prime\prime}}} =
a_1 b_1 + \omega a_2 b_2 + \omega^2 a_3 b_3,
\label{33:1primeprime}
\\
&&(a \otimes b)_{\mathbf{3}_s} =
(a_2 b_3 + a_3 b_2, a_3 b_1 + a_1 b_3, a_1 b_2 + a_2 b_1),
\label{33:3s}
\nonumber\\
&&(a \otimes b)_{\mathbf{3}_a} =
(a_2 b_3 - a_3 b_2, a_3 b_1 - a_1 b_3, a_1 b_2 - a_2 b_1),
\nonumber\label{33:3a}
\end{eqnarray}
where $\omega = e^{2 i \pi/3}$, and $s, a$ stand for the symmetric and
anti-symmetric triplet components, respectively.

We will also need the product of three triplets, $a$, $b$, and $c=(c_1, c_2, c_3)$:
\begin{eqnarray}
(a \otimes b \otimes c)_s
&=&
a_1 (b_2 c_3 + b_3 c_2)
+
a_2 (b_3 c_1 + b_1 c_3)\nonumber\\
& & +
a_3 (b_1 c_2 + b_2 c_1),
\label{333:s}
\nonumber\\
(a \otimes b \otimes c)_a
&=&
a_1 (b_2 c_3 - b_3 c_2)
+
a_2 (b_3 c_1 - b_1 c_3)\nonumber\\
& & +
a_3 (b_1 c_2 - b_2 c_1).
\label{333:a}
\end{eqnarray}

We are now ready to construct the Yukawa matrices for the various cases.
We have built a program to test all possibilities automatically. As a first
example, let us consider the case $\Phi \sim \mathbf{3}$, $(\overline{Q}_{L1},
\overline{Q}_{L2}, \overline{Q}_{L3}) \sim (\mathbf{1}, \mathbf{1}, \mathbf{1^\prime})$,
$d_R \sim \mathbf{3}$, and $u_R \sim \mathbf{3}$. We start with the down sector.
Since $\overline{Q}_{L1}$ is in the $\mathbf{1}$ representation, it must couple
to the $(\Phi \otimes d_R)_\mathbf{1}$ combination obtained from Eq.~\eqref{33:1}.
The same is true for $\overline{Q}_{L2}$, with an independent coefficient.
This leads to the Yukawa terms
\begin{eqnarray}
&&
\alpha_1\,
\overline{Q}_{L1}
\left[
\Phi_1 d_{R1} +  \Phi_2 d_{R2}
+ \Phi_3 d_{R3}
\right]
\nonumber\\
&&
+\,
\alpha_2\,
\overline{Q}_{L2}
\left[
\Phi_1 d_{R1} +  \Phi_2 d_{R2}
+ \Phi_3 d_{R3}
\right].
\end{eqnarray}
Once the fields $\Phi_i$ are substituted by their vevs $v_i$, these terms give
the first and second row of the down-type quark mass matrix, $M_d$, respectively.
Since $\overline{Q}_{L3}$ is in the $\mathbf{1^\prime}$ representation, we can
only obtain a singlet with the $\mathbf{1^{\prime\prime}}$ combination
$(\Phi \otimes d_R)_\mathbf{1^{\prime\prime}}$ in Eq.~\eqref{33:1primeprime}.
This leads to a term
\begin{equation}
\alpha_3\,
\overline{Q}_{L3}
\left[
\Phi_1 d_{R1} + \omega \Phi_2 d_{R2}
+ \omega^2 \Phi_3 d_{R3}
\right],
\end{equation}
which will fill the third row of $M_d$. Thus, the down-type quark mass matrix
reads
\begin{equation}
M_d =
\left(
\begin{array}{ccc}
\alpha_1 v_1 & \alpha_1 v_2 & \alpha_1 v_3\\
\alpha_2 v_1 & \alpha_2 v_2 & \alpha_2 v_3\\
\alpha_3 v_1 & \omega \alpha_3 v_2 & \omega^2 \alpha_3 v_3
\end{array}
\right),
\label{Yd}
\end{equation}
with arbitrary complex constants $\alpha_i$.

Recalling that the up-quark Yukawa terms involve the combinations
$\overline{Q}_L \tilde{\Phi}\, u_R$, a similar analysis of the up-type quark
sector yields
\begin{equation}
M_u =
\left(
\begin{array}{ccc}
\beta_1 v_1^\ast & \beta_1 v_2^\ast & \beta_1 v_3^\ast\\
\beta_2 v_1^\ast & \beta_2 v_2^\ast & \beta_2 v_3^\ast\\
\beta_3 v_1^\ast & \omega \beta_3 v_2^\ast & \omega^2 \beta_3 v_3^\ast
\end{array}
\right),
\label{Yu}
\end{equation}
where $\beta_i$ are arbitrary complex constants.

In order to find the most relevant features of the quark sector, we define
the Hermitian matrices
\begin{equation}
H_d = M_d M_d^\dagger,\quad
H_u = M_u M_u^\dagger,
\label{HdHu}
\end{equation}
whose eigenvalues coincide with the squared masses in each quark sector.
Moreover, the CKM CP-violating phase is proportional to the determinant~\cite{Ja}
\begin{equation}
J = \textrm{Det} (H_d H_u - H_u H_d).
\end{equation}
We must now substitute $(v_1, v_2, v_3)$ by each of the possible vev alignments
in Eq.~\eqref{vev:A4}, including all possible permutations, and study the
properties of $H_d$, $H_u$, and $J$. As an example, consider the possibility
that $(v_1, v_2, v_3) = v (1, e^{i \alpha}, 0)$, for any phase $\alpha$. Then
\begin{eqnarray}
M_d
&=&
v \left(
\begin{array}{ccc}
\alpha_1  & \alpha_1 e^{i \alpha} & 0\\
\alpha_2  & \alpha_2 e^{i \alpha} & 0\\
\alpha_3  & \omega \alpha_3 e^{i \alpha} & 0
\end{array}
\right),
\label{Yd_case}
\\
M_u
&=&
v
\left(
\begin{array}{ccc}
\beta_1  & \beta_1 e^{-i \alpha} & 0\\
\beta_2  & \beta_2 e^{-i \alpha} & 0\\
\beta_3  & \omega \beta_3 e^{-i \alpha} & 0
\end{array}
\right).
\label{Yu_case}
\end{eqnarray}
As a result, we predict one massless quark with charge $-1/3$ and one massless
quark with charge $2/3$, contrary to experimental evidence. It is interesting
to note that, in this case, $H_d$ and $H_u$ do not depend on $\alpha$ but,
nevertheless, $J \neq 0$. This means that the model predicts one massless quark
in each charge sector but displays explicit CP violation in the CKM
matrix~\footnote{One could envisage a more complicated setup where the light
quark masses appear radiatively.}.

As a second example, let us consider the case $\Phi \sim \mathbf{3}$,
$(\overline{Q}_{L1}, \overline{Q}_{L2}, \overline{Q}_{L3}) \sim (\mathbf{1},
\mathbf{1^\prime}, \mathbf{1^{\prime \prime}})$, $d_R \sim \mathbf{3}$, and
$u_R \sim \mathbf{3}$. We find
\begin{eqnarray}
M_d
&=&
\left(
\begin{array}{ccc}
\alpha_1 v_1 & \alpha_1 v_2 & \alpha_1 v_3\\
\alpha_2 v_1 & \omega \alpha_2 v_2 & \omega^2 \alpha_2 v_3\\
\alpha_3 v_1 & \omega^2 \alpha_3 v_2 & \omega \alpha_3 v_3
\end{array}
\right),
\label{Yd_2}
\\
M_u
&=&
\left(
\begin{array}{ccc}
\beta_1 v_1^\ast & \beta_1 v_2^\ast & \beta_1 v_3^\ast\\
\beta_2 v_1^\ast & \omega \beta_2 v_2^\ast & \omega^2 \beta_2 v_3^\ast\\
\beta_3 v_1^\ast & \omega^2 \beta_3 v_2^\ast & \omega \beta_3 v_3^\ast
\end{array}
\right).
\label{Yu_2}
\end{eqnarray}
For the vev alignments $v(1,1,1)$ and $v(\pm 1, \eta, \eta^\ast)$ of
Eq.~\eqref{vev:A4}, this leads to
\begin{eqnarray}
H_d
&=&
3 v^2
\left(
\begin{array}{ccc}
|\alpha_1|^2  & 0 & 0\\
0 & |\alpha_2|^2 & 0\\
0 & 0 & |\alpha_3|^2
\end{array}
\right),
\label{Hd_2}
\\
H_u
&=&
3 v^2
\left(
\begin{array}{ccc}
|\beta_1|^2  & 0 & 0\\
0 & |\beta_2|^2 & 0\\
0 & 0 & |\beta_3|^2
\end{array}
\right),
\label{Hu_2}
\end{eqnarray}
meaning that, in these cases, all quark masses are non-vanishing and non-degenerate.
However, we find a diagonal CKM matrix and no CP-violation, in blatant contradiction
with experiment.

The particular case where $\overline{Q}_L$, $u_R$, and $d_R$ (in addition to $\Phi$)
are all in a triplet representation of $A_4$ has been considered in
Refs.~\cite{Wyler:1979fe,Branco:1980ax} for the first three vevs given in
Eq.~\eqref{vev:A4}. Ref.~\cite{Wyler:1979fe} solves the problem by adding a fourth
scalar as a singlet of $A_4$; Ref.~\cite{Branco:1980ax} considers symmetry
breaking in stages.

Having gone through all cases in Table~\ref{tabA4:rep} and all possible vev
alignments in Eq.~\eqref{vev:A4} (including permutations), we find that in all situations one obtains either massless quarks or a vanishing CKM phase.

In Table~\ref{TabA4full} we present, for each choice of representations and for each vev alignment given in Eq.~\eqref{vev:A4}, the different quark mass spectra and the number of CKM mixing angles not predicted by the discrete symmetry, i.e the number of parameter-dependent mixing angles (PDMA).

\begin{table}[H]
\begin{center}
\begin{tabular}{|c|cccccccc||ccccl|}
\hline
vev&& $Q_L$ & & $u_R$ & & $d_R$ & & & & &$\begin{array}{c}\text{Number of}\\\text{PDMA}\end{array}$& &$\begin{array}{l}\text{Mass}\\\text{spectrum}\end{array}$\\
\hline\hline
\multirow{9}{*}{\rotatebox{90}{$(1,0,0)$}}&& $\mathbf{3}$ & & $\mathbf{3}$ & & $\mathbf{3}$ & &&  & & $0$ & & $(0,m_{u,d},m_{u,d}^\prime)$\\
\cline{2-14}
&& $\mathbf{3}$ & & $\mathbf{3}$ & & $\mathbf{s}$ & &  && & $0$ & & $\begin{array}{l}(0,m_{u},m_{u}^\prime)\\(0,0,m_{d})\end{array}$\\
\cline{2-14}
&& $\mathbf{3}$ & & $\mathbf{s}$  & & $\mathbf{3}$ & &&  & & $0$ & &  $\begin{array}{l}(0,0,m_{u})\\(0,m_{d},m_{d}^\prime)\end{array}$ \\
\cline{2-14}
&& $\mathbf{3}$ & & $\mathbf{s}$  & & $\mathbf{s}$ & & & & & $0$ & & $(0,0,m_{u,d})$\\
\cline{2-14}
&& $\mathbf{s}$ & & $\mathbf{3}$  & & $\mathbf{3}$ & & & & & $2$ & & $(0,0,m_{u,d})$\\
\hline\hline
\multirow{11}{*}{\rotatebox{90}{$(1,1,1)\,,\,(\pm1,\eta,\eta^\ast)$}}&& $\mathbf{3}$ & & $\mathbf{3}$ & & $\mathbf{3}$ & &&  & & $0$ & &
$(m_{u,d},m_{u,d}^\prime,m_{u,d}^{\prime\prime})$ \\
\cline{2-14}
&& $\mathbf{3}$ & & $\mathbf{3}$ & & $\mathbf{s}$ & &  && & $0$ & & $\begin{array}{l}(m_{u},m_{u}^\prime,m_{u}^{\prime\prime})\\(\times,\times,m_d)\end{array}$\\
\cline{2-14}
&& $\mathbf{3}$ & & $\mathbf{s}$  & & $\mathbf{3}$ & &&  & & $0$ & &  $\begin{array}{l}(\times,\times,m_u)\\(m_{d},m_{d}^\prime,m_{d}^{\prime\prime})\end{array}$ \\
\cline{2-14}
&& $\mathbf{3}$ & & $\mathbf{s}$  & & $\mathbf{s}$ & & & & & $0$ & & $(\times,\times,m_{u,d})$ \\
\cline{2-14}
&& $\mathbf{s}$ & & $\mathbf{3}$  & & $\mathbf{3}$ & & & & & $\begin{array}{l}0\\ 1\\ 2\end{array}$ & &
$\begin{array}{l}(m_{u,d},m_{u,d}^\prime,m_{u,d}^{\prime\prime})\\(0,m_{u,d},m_{u,d}^\prime)\\
(0,0,m_{u,d})\end{array}$ \\
\hline
\hline
\multirow{10}{*}{
\rotatebox{90}{$(
1,e^{i\alpha},0)$}}&& $\mathbf{3}$ & & $\mathbf{3}$ & & $\mathbf{3}$ & &&  & & $1$ & & $(0,m_{u,d},m_{u,d})$\\
\cline{2-14}
&& $\mathbf{3}$ & & $\mathbf{3}$ & & $\mathbf{s}$ & &  && & $1$ & & $\begin{array}{l}(0,m_u,m_u)\\(0,\times,m_d)\end{array}$ \\
\cline{2-14}
&& $\mathbf{3}$ & & $\mathbf{s}$  & & $\mathbf{3}$ & &&  & & $1$ & &  $\begin{array}{l}(0,\times,m_u)\\(0,m_d,m_d)\end{array}$ \\
\cline{2-14}
&& $\mathbf{3}$ & & $\mathbf{s}$  & & $\mathbf{s}$ & & & & & $1$ & & $(0,\times,m_{u,d})$ \\
\cline{2-14}
&& $\mathbf{s}$ & & $\mathbf{3}$  & & $\mathbf{3}$ & & & & & $\begin{array}{l}3\\ 2\end{array}$ & &$\begin{array}{l}(0,m_{u,d},m_{u,d}^\prime)
\\(0,0,m_{u,d})\end{array}$\\
\hline
\end{tabular}
\caption{Quark mass spectra and number of arbitrary CKM parameter-dependent
mixing angles (PDMA) in the $A_4$ case. The symbol $\times$ stands for $0$ or $m_i\neq0$; $\mathbf{s}$ stands for $\mathbf{1},\mathbf{1^\prime}$ or
$\mathbf{1^{\prime \prime}}$. \label{TabA4full}}
\end{center}
\end{table}

Requiring non-vanishing quarks by itself, restricts the representations of
$\{ Q_L; u_R; d_R \}$ to the five possibilities $\{ \mathbf{s}; \mathbf{3};
\mathbf{3} \}$, $\{ \mathbf{3}; \mathbf{s}; \mathbf{s} \}$, $\{ \mathbf{3};
\mathbf{s}; \mathbf{3} \}$, $\{ \mathbf{3}; \mathbf{3}; \mathbf{s} \}$, and
$\{ \mathbf{3}; \mathbf{3}; \mathbf{3} \}$, where $\mathbf{s}$ stands for
$(\mathbf{1},\mathbf{1^\prime},\mathbf{1^{\prime \prime}})$, with the vevs
restricted to $v(1,1,1)$ or $v(\pm 1, \eta, \eta^\ast)$. In all these special
cases, the CKM matrix equals the unit matrix. Thus, it is not possible to extend
the $A_4$ symmetry to the quark sector, with only three generations of quarks and
the three scalar fields in a triplet of $A_4$.

It is conceivable that this problem can be evaded by adding quark generations.
More commonly, one considers other representations for the three scalar fields
and/or one adds extra scalars to the theory in other representations of $A_4$.
But, in such cases one \textit{must} prove that the local minimum does indeed
correspond to a global minimum. One can see from the treatment of $A_4$ that
this endeavor is far from trivial~\cite{ivanov_min}.

\section{\label{sec:S4}The $S_4$ case}

$S_4$ is the group of all permutations of four objects. It has 24 elements divided
into five irreducible representations: two singlets $\mathbf{1_1}, \ \mathbf{1_2}$,
one doublet $\ \mathbf{2}$ and two triplets $\ \mathbf{3_1}, \ \mathbf{3_2}$. The
multiplication rules are:
\begin{eqnarray}\label{S4rules}
\mathbf{1_1} \otimes \textrm{any}
&=&
\textrm{any},
\nonumber\\
\mathbf{1_2} \otimes \mathbf{1_2}
&=&
\mathbf{1_1},
\nonumber\\
\mathbf{1_2} \otimes \mathbf{2}
&=&
\mathbf{2},
\nonumber\\
\mathbf{1_2} \otimes \mathbf{3_1}
&=&
\mathbf{3_2},
\nonumber\\
\mathbf{1_2} \otimes \mathbf{3_2}
&=&
\mathbf{3_1},
\nonumber\\
\mathbf{2} \otimes \mathbf{2}
&=&
\mathbf{1_1} \oplus \mathbf{1_2} \oplus \mathbf{2},
\\
\mathbf{2} \otimes \mathbf{3_1}
&=&
\mathbf{3_1} \oplus \mathbf{3_2} ,
\nonumber\\
\mathbf{2} \otimes \mathbf{3_2}
&=&
\mathbf{3_1} \oplus \mathbf{3_2},
\nonumber\\
\mathbf{3_1} \otimes \mathbf{3_1}
&=&
\mathbf{1_1} \oplus \mathbf{2} \oplus \mathbf{3_1} \oplus \mathbf{3_2} ,
\nonumber\\
\mathbf{3_1} \otimes \mathbf{3_2}
&=&
\mathbf{1_2} \oplus \mathbf{2} \oplus \mathbf{3_1} \oplus \mathbf{3_2} ,
\nonumber\\
\mathbf{3_2} \otimes \mathbf{3_2}
&=&
\mathbf{1_1} \oplus \mathbf{2} \oplus \mathbf{3_1} \oplus \mathbf{3_2}.
\nonumber
\end{eqnarray}
Since $A_4$ is a subgroup of $S_4$, this case will have at least the same unphysical restrictions. Yet, for model building, it is useful to go through the analysis in detail, uncovering the specific constraints that should be corrected when enlarging the model.

Let us start by assuming that the three Higgs doublets are in the representation
$\mathbf{3_1}$. By looking at Eqs.~\eqref{S4rules}, we see that the product of
left-handed and right-handed fermions must also be in a $\mathbf{3_1}$
representation (or else, the Yukawa Lagrangian would not be in the invariant
$\mathbf{1_1}$ representation). The possibilities for the representations of
the up and down right-handed quarks are listed in Table~\ref{tabS4:rep-Phi3_1-QL3},
when $Q_L$ is in a triplet representation.
\begin{table}[h!]
\begin{center}
\begin{tabular}{|ccccccc||ccccccc|}
\hline
& $Q_L$ & & $u_R$ & & $d_R$ & &
& $Q_L$ & & $u_R$ & & $d_R$ & \\
\hline
 & $\mathbf{3_1}$ & &
$\mathbf{1_1}$, $\mathbf{1_1}$, $\mathbf{1_1}$  & &
$\mathbf{1_1}$, $\mathbf{1_1}$, $\mathbf{1_1}$ & &
 & $\mathbf{3_2}$ & &
$\mathbf{1_2}$, $\mathbf{1_2}$, $\mathbf{1_2}$  & &
$\mathbf{1_2}$, $\mathbf{1_2}$, $\mathbf{1_2}$& \\
 & & &
$\mathbf{1_1}$, $\mathbf{1_1}$, $\mathbf{1_1}$ & &
$\mathbf{2}$, $\mathbf{1_1}$ &
 & & & &
$\mathbf{1_2}$, $\mathbf{1_2}$, $\mathbf{1_2}$  & &
$\mathbf{2}$, $\mathbf{1_2}$ & \\
  & & &
$\mathbf{1_1}$, $\mathbf{1_1}$, $\mathbf{1_1}$  & &
$\mathbf{3_1}$ &
  & & & &
$\mathbf{1_2}$, $\mathbf{1_2}$, $\mathbf{1_2}$ & &
$\mathbf{3_1}$ & \\
  & & &
$\mathbf{1_1}$, $\mathbf{1_1}$, $\mathbf{1_1}$  & &
$\mathbf{3_2}$ &
  & & & &
$\mathbf{1_2}$, $\mathbf{1_2}$, $\mathbf{1_2}$  & &
$\mathbf{3_2}$ & \\
 & & &
$\mathbf{2}$, $\mathbf{1_1}$ & &
$\mathbf{1_1}$, $\mathbf{1_1}$, $\mathbf{1_1}$ &
 & & & &
$\mathbf{2}$, $\mathbf{1_2}$  & &
$\mathbf{1_2}$, $\mathbf{1_2}$, $\mathbf{1_2}$ & \\
 & & &
$\mathbf{2}$, $\mathbf{1_1}$ & &
$\mathbf{2}$, $\mathbf{1_1}$ &
 & & & &
$\mathbf{2}$, $\mathbf{1_2}$  & &
$\mathbf{2}$, $\mathbf{1_2}$ & \\
  & & &
$\mathbf{2}$, $\mathbf{1_1}$  & &
$\mathbf{3_1}$ &
  & & & &
$\mathbf{2}$, $\mathbf{1_2}$  & &
$\mathbf{3_1}$ & \\
 & & &
$\mathbf{2}$, $\mathbf{1_1}$  & &
$\mathbf{3_2}$ &
 & & & &
$\mathbf{2}$, $\mathbf{1_2}$  & &
$\mathbf{3_2}$ & \\
 & & &
$\mathbf{3_1}$  & &
$\mathbf{1_1}$, $\mathbf{1_1}$, $\mathbf{1_1}$ &
 & & & &
$\mathbf{3_1}$  & &
$\mathbf{1_2}$, $\mathbf{1_2}$, $\mathbf{1_2}$ & \\
 & & &
$\mathbf{3_1}$  & &
$\mathbf{2}$, $\mathbf{1_1}$&
& & & &
$\mathbf{3_1}$  & &
$\mathbf{2}$, $\mathbf{1_2}$ & \\
 & & &
$\mathbf{3_1}$  & &
$\mathbf{3_1}$ &
 & & & &
$\mathbf{3_1}$  & &
$\mathbf{3_1}$ & \\
 & & &
$\mathbf{3_1}$  & &
$\mathbf{3_2}$ &
 & & & &
$\mathbf{3_1}$  & &
$\mathbf{3_2}$ & \\
 & & &
$\mathbf{3_2}$  & &
$\mathbf{1_1}$, $\mathbf{1_1}$, $\mathbf{1_1}$ &
 & & & &
$\mathbf{3_2}$  & &
$\mathbf{1_2}$, $\mathbf{1_2}$, $\mathbf{1_2}$ & \\
 & & &
$\mathbf{3_2}$  & &
$\mathbf{2}$, $\mathbf{1_1}$ &
 & & & &
$\mathbf{3_2}$  & &
$\mathbf{2}$, $\mathbf{1_2}$ & \\
 & & &
$\mathbf{3_2}$  & &
$\mathbf{3_1}$ &
 & & & &
$\mathbf{3_2}$  & &
$\mathbf{3_1}$ & \\
 & & &
$\mathbf{3_2}$  & &
$\mathbf{3_2}$ &
 & & & &
$\mathbf{3_2}$  & &
$\mathbf{3_2}$ & \\
\hline
\end{tabular}
\end{center}
\caption{Possible representations of $u_R$ and $d_R$ when the three Higgs
doublets are in a $\mathbf{3_1}$ representation and all $Q_L$ are in a triplet
representation $\mathbf{3_1}$ or $\mathbf{3_2}$.
\label{tabS4:rep-Phi3_1-QL3}}
\end{table}

When two of the $Q_L$ are in the doublet $\mathbf{2}$ representation, the
possibilities are $(Q_L, u_R, d_R) \sim (\mathbf{2}, \mathbf{3_1},
\mathbf{3_1})$, $(\mathbf{2}, \mathbf{3_1}, \mathbf{3_2})$, $(\mathbf{2},
\mathbf{3_2}, \mathbf{3_1})$, or $(\mathbf{2}, \mathbf{3_2}, \mathbf{3_2})$.
Similarly, when one of the $Q_L$ is in a singlet representation, there are only
two possibilities: either $(Q_L, u_R, d_R) \sim ( \mathbf{1_1}, \mathbf{3_1},
\mathbf{3_1})$, or  $(Q_L, u_R, d_R) \sim (\mathbf{1_2}, \mathbf{3_2},
\mathbf{3_2})$. But, in this case, the third $Q_L$ field must be in a singlet
representation that yields a Yukawa Lagrangian in the singlet representation.
Otherwise, the mass matrix would have a row of zeros, and there would be a
massless quark. As a result, when two of the $Q_L$ are in the doublet
$\mathbf{2}$ representation, the only viable possibilities for $u_R$ and $d_R$
are the ones listed in Table~\ref{tabS4:rep-Phi3_1-QL2}.
\begin{table}[h!]
\begin{center}
\begin{tabular}{|ccccccc|}
\hline
& $Q_L$ & & $u_R$ & & $d_R$ & \\
\hline
& $\mathbf{2}$, $\mathbf{1_1}$ & &
$\mathbf{3_1}$  & &
$\mathbf{3_1}$ & \\
& $\mathbf{2}$, $\mathbf{1_2}$ & &
$\mathbf{3_2}$  & &
$\mathbf{3_2}$ & \\
\hline
\end{tabular}
\end{center}
\caption{Possible representations of $u_R$ and $d_R$ when the three Higgs
doublets are in a $\mathbf{3_1}$ representation and two of the $Q_L$ are in
the doublet representation $\mathbf{2}$.
\label{tabS4:rep-Phi3_1-QL2}}
\end{table}

Finally, requiring that there are no massless quarks, when all the $Q_L$ are
in a singlet representation, the possibilities for $u_R$ and $d_R$ are listed
in Table~\ref{tabS4:rep-Phi3_1-QL1}.
\begin{table}[h!]
\begin{center}
\begin{tabular}{|ccccccc|}
\hline
& $Q_L$ & & $u_R$ & & $d_R$ & \\
\hline
& $\mathbf{1_1}$, $\mathbf{1_1}$, $\mathbf{1_1}$  & &
$\mathbf{3_1}$  & &
$\mathbf{3_1}$ & \\
& $\mathbf{1_2}$, $\mathbf{1_2}$, $\mathbf{1_2}$ & &
$\mathbf{3_2}$  & &
$\mathbf{3_2}$ & \\
\hline
\end{tabular}
\end{center}
\caption{Possible representations of $u_R$ and $d_R$ when the three Higgs
doublets are in a $\mathbf{3_1}$ representation and all $Q_L$ are in a singlet
representation $\mathbf{1_1}$ or $\mathbf{1_2}$.
\label{tabS4:rep-Phi3_1-QL1}}
\end{table}

A suitable basis for the $\mathbf{3_1}$ representation of $S_4$, consistent
with the notation of Ref.~\cite{ivanov_min}, can be found in
Ref.~\cite{lavoura_varzielas}:
\begin{equation}
F_3 =
\left(
\begin{array}{ccc}
1 & 0 & 0\\
0 & 0 & -1\\
0 & -1 & 0
\end{array}
\right),
\quad
G_3 =
\left(
\begin{array}{ccc}
0 & 1 & 0\\
0 & 0 & 1\\
1 & 0 & 0
\end{array}
\right).
\label{F3&G3}
\end{equation}
Notice that $G_3$ coincides with $T$ in Eq.~\eqref{S&T}. Imposing $F_3$ and
$G_3$ on the 3HDM potential we recover Eq.~\eqref{masterV}, with
$\Lambda_4 = 0$. The $\mathbf{3_2}$ representation of $S_4$ can be identified
with the matrices $-F_3$ and $G_3$. These matrices satisfy
$F_3^2=G_3^3=(F_3 G_3)^4 = 1$, showing that they indeed generate the group $S_4$.
 As for the explicit form of the tensor products, we will use the Appendix of
 Ref.~\cite{lavoura_varzielas}. For example, the product of two $\mathbf{3_1}$
 triplets, $a=(a_1, a_2, a_3)$ and $b=(b_1, b_2, b_3)$, gives
\begin{eqnarray}\label{S4prod}
&&(a \otimes b)_{\mathbf{1_1}} =
a_1 b_1 + a_2 b_2 + a_3 b_3,
\nonumber\\
&&(a \otimes b)_{\mathbf{2}} =
(a_1 b_1 + \omega a_2 b_2 + \omega^2 a_3 b_3,\\
&&\qquad \qquad \quad a_1 b_1 + \omega^2 a_2 b_2 + \omega a_3 b_3),
\nonumber\\
&&(a \otimes b)_{\mathbf{3_1}} =
(a_2 b_3 + a_3 b_2, a_3 b_1 + a_1 b_3, a_1 b_2 + a_2 b_1),
\nonumber\\
&&(a \otimes b)_{\mathbf{3_2}} =
(a_2 b_3 - a_3 b_2, a_3 b_1 - a_1 b_3, a_1 b_2 - a_2 b_1).\nonumber
\end{eqnarray}

For illustration, let us consider the case $\Phi \sim \mathbf{3_1}$,
$(\overline{Q}_{L1}, \overline{Q}_{L2}) \sim \mathbf{2}$, $\overline{Q}_{L3}
\sim \mathbf{1_1}$, $d_R \sim \mathbf{3_1}$, and $u_R \sim \mathbf{3_1}$. We
start with the down sector. The fact that $(\overline{Q}_{L1}, \overline{Q}_{L2})$
is in the doublet representation $\mathbf{2}$, means that we must pick up the
doublet combination $(\Phi \otimes d_R)_\mathbf{2}$ obtained from
Eq.~\eqref{S4prod}, leading to
\begin{eqnarray}
&&
\alpha_1\,
\overline{Q}_{L1}
\left[
\Phi_1 d_{R1} +  \omega \Phi_2 d_{R2}
+ \omega^2 \Phi_3 d_{R3}
\right]
\nonumber\\
&&
+\,\alpha_1\,\overline{Q}_{L2}
\left[
\Phi_1 d_{R1} +  \omega^2 \Phi_2 d_{R2}
+ \omega \Phi_3 d_{R3}
\right].
\end{eqnarray}
On the other hand, $\overline{Q}_{L3} \sim \mathbf{1_1}$ couples to
$(\Phi \otimes d_R)_\mathbf{1_1}$ in Eq.~\eqref{S4prod}, yielding
\begin{equation}
\alpha_2\,
\overline{Q}_{L3}
\left[
\Phi_1 d_{R1} +  d_{R2} + \Phi_3 d_{R3}
\right].
\end{equation}
Hence,
\begin{equation}
M_d =
\left(
\begin{array}{ccc}
\alpha_1 v_1 & \omega \alpha_1 v_2 & \omega^2 \alpha_1 v_3\\
\alpha_1 v_1 & \omega^2 \alpha_1 v_2 & \omega \alpha_1 v_3\\
\alpha_2 v_1 & \alpha_2 v_2 & \alpha_2 v_3
\end{array}
\right).
\label{Yd_S4}
\end{equation}
Similarly,
\begin{equation}
M_u =
\left(
\begin{array}{ccc}
\beta_1 v_1^\ast & \omega \beta_1 v_2^\ast & \omega^2 \beta_1 v_3^\ast\\
\beta_1 v_1^\ast & \omega^2 \beta_1 v_2^\ast & \omega \beta_1 v_3^\ast\\
\beta_2 v_1^\ast & \beta_2 v_2^\ast & \beta_2 v_3^\ast
\end{array}
\right).
\label{Yu_S4}
\end{equation}
The predictions for the physical observables should now be found for all the
possible global minima presented in Eq.~\eqref{vev:S4}. Let us test the case
with the vev alignment $v (1,1,1)$. We find
\begin{eqnarray}
H_d
&=&
3 v^2
\left(
\begin{array}{ccc}
|\alpha_1|^2 & 0 & 0\\
0 & |\alpha_1|^2 & 0\\
0 & 0 & |\alpha_2|^2
\end{array}
\right),
\nonumber\\
H_u
&=&
3 v^2
\left(
\begin{array}{ccc}
|\beta_1|^2 & 0 & 0\\
0 & |\beta_1|^2 & 0\\
0 & 0 & |\beta_2|^2
\end{array}
\right).
\end{eqnarray}

Although this case does not exhibit massless quarks, it has a pair of degenerate quarks in each sector, the CKM is the unit matrix and, of course, there is no CP violation.

The analysis for $\Phi \sim \mathbf{3_2}$ leads to a new set of cases obtained trivially from Tables~\ref{tabS4:rep-Phi3_1-QL3}, \ref{tabS4:rep-Phi3_1-QL2}, and \ref{tabS4:rep-Phi3_1-QL1}, by noting that $\mathbf{3_2} = \mathbf{3_1} \otimes \mathbf{1_2}$. As we did for $A_4$, we have also built a program to test all $S_4$ possibilities automatically. In all cases, there is no CP violation in the CKM matrix ($J=0$) and, in the absence of massless quarks, there will always be one pair of degenerate quarks in each sector. The restrictions on the physical parameters obtained for each choice of representations and for each vev alignment in Eq.~\eqref{vev:S4}, can be found in Tables~\ref{TabS4full1}-\ref{TabS4full3}. This may help model builders in identifying what features need to be corrected when adding extra fields to the theory.

\begin{table}[t]
\begin{center}
\begin{tabular}{|c|ccc|cccc||cl|}
\hline
vev&& $Q_L$ & & $u_R$ & & $d_R$ & & $\begin{array}{c}\text{Number of}\\\text{PDMA}\end{array}$& $\begin{array}{l}\text{Mass}\\\text{spectrum}\end{array}$\\
\hline\hline
\multirow{17}{*}{\rotatebox{90}{$(1,0,0)$}}&& $\mathbf{3_1}$ & & $\mathbf{1_1}$ & & $\mathbf{1_1}$  & & $0$ & $(0,0,m_{u,d})$\\
\cline{5-10}
&&  & & $\mathbf{1_1}$ & & $\mathbf{2,\,1_1}$ & &  $0$ & $(0,0,m_{u,d})$\\
\cline{5-10}
&&  & & $\mathbf{1_1}$  & & $\mathbf{3_i}$ & & $0$ &  $\begin{array}{l}(0,0,m_u)\\(0,m_d,m_d)\end{array}$ \\
\cline{5-10}
&&  & & $\mathbf{2,\,1_1}$  & & $\mathbf{1_1}$ & &  $0$ &  $(0,0,m_{u,d})$\\
\cline{5-10}
&&  & & $\mathbf{2,\,1_1}$  & & $\mathbf{2,\,1_1}$ & & $0$ &  $(0,0,m_{u,d})$\\
\cline{5-10}
&&  & & $\mathbf{2,\,1_1}$  & & $\mathbf{3_i}$ & & $0$ &$\begin{array}{l}(0,0,m_u)\\(0,m_d,m_d)\end{array}$\\
\cline{5-10}
&&  & & $\mathbf{3_i}$  & & $\mathbf{1_1}$ & & $0$ &  $\begin{array}{l}(0,m_u,m_u)\\(0,0,m_d)\end{array}$\\
\cline{5-10}
&&  & & $\mathbf{3_i}$  & & $\mathbf{2,\,1_1}$ & & $0$ & $\begin{array}{l}(0,m_u,m_u)\\(0,0,m_d)\end{array}$\\
\cline{5-10}
&&  & & $\mathbf{3_i}$  & & $\mathbf{3_j}$ & & $0$ & $(0,m_{u,d},m_{u,d})$\\
\cline{2-10}
&& $\mathbf{2,\,1_i}$  & & $\mathbf{3_i}$  & & $\mathbf{3_i}$ & & $1$ &$(0,0,m_{u,d})$\\
\cline{2-10}
&& $\mathbf{1_i}$  & & $\mathbf{3_i}$  & & $\mathbf{3_i}$ & & $2$ & $(0,0,m_{u,d})$\\
\hline
\end{tabular}
\caption{Quark mass spectra and number of arbitrary CKM parameter-dependent
mixing angles (PDMA) in the $S_4$ case, for the vev $v\,(1,0,0)$. In all cases, $\Phi \sim \mathbf{3_1}$.\label{TabS4full1}}
\end{center}
\end{table}

\begin{table}[H]
\begin{center}
\begin{tabular}{|c|ccc|cccc||cl|}
\hline
vev&& $Q_L$ & & $u_R$ & & $d_R$ & &  $\begin{array}{c}\text{Number of}\\\text{PDMA}\end{array}$& $\begin{array}{l}\text{Mass}\\\text{spectrum}\end{array}$\\
\hline\hline
\multirow{20}{*}{\rotatebox{90}{$(1,1,1)\,,\,(\pm1,\eta,\eta^\ast)$}}&& $\mathbf{3_1}$ & & $\mathbf{1_1}$ & & $\mathbf{1_1}$ & &  $0$ &  $(0,0,m_{u,d})$\\
 \cline{5-10}
&&  & & $\mathbf{1_1}$ & & $\mathbf{2,\,1_1}$ & &   $0$ &  $\begin{array}{l}(0,0,m_u)\\(m_d,m_d,m_d^\prime)\end{array}$\\
 \cline{5-10}
&&  & & $\mathbf{1_1}$  & & $\mathbf{3_i}$ & &   $0$ &   $\begin{array}{l}(0,0,m_u)\\(m_d,m_d,2 m_d\,\delta_{1i})\end{array}$ \\
\cline{5-10}
&&  & & $\mathbf{2,\,1_1}$  & & $\mathbf{1_1}$ & &   $0$ &  $\begin{array}{l}(m_u,m_u,m_u^\prime)\\(0,0,m_d)\end{array}$\\
\cline{5-10}
&&  & & $\mathbf{2,\,1_1}$  & & $\mathbf{2,\,1_1}$ & &   $0$ &  $(m_{u,d},m_{u,d},m_{u,d}^\prime)$\\
\cline{5-10}
&&  & & $\mathbf{2,\,1_1}$  & & $\mathbf{3_i}$ & &   $0$ &  $\begin{array}{l}(m_u,m_u,m_u^\prime)\\(m_d,m_d,2 m_d\,\delta_{1i})\end{array}$\\
\cline{5-10}
&&  & & $\mathbf{3_i}$  & & $\mathbf{1_1}$ & &   $0$ &  $\begin{array}{l}(m_u,m_u,2m_u\,\delta_{1i})\\(0,0,m_d)\end{array}$\\
\cline{5-10}
&&  & & $\mathbf{3_i}$  & & $\mathbf{2,\,1_1}$ & &   $0$ &  $\begin{array}{l}(m_u,m_u,2m_u\,\delta_{1i})\\(m_d,m_d,m_d^\prime)\end{array}$\\
\cline{5-10}
&&  & & $\mathbf{3_i}$  & & $\mathbf{3_j}$ & &   $0$ &  $\begin{array}{l}(m_u,m_u,2m_u\,\delta_{1i})\\(m_d,m_d,2m_d\,\delta_{1j})\end{array}$\\
\cline{2-10}
&& $\mathbf{2,\,1_i}$  & & $\mathbf{3_i}$  & & $\mathbf{3_i}$ & &   $0$ &  $(m_{u,d},m_{u,d},m_{u,d}^\prime)$\\
\cline{2-10}
&& $\mathbf{1_i}$  & & $\mathbf{3_i}$  & & $\mathbf{3_i}$ & &  $2$ &  $(0,0,m_{u,d})$\\
\hline
\end{tabular}
\caption{As in Table~\ref{TabS4full1}; for the vev $v\,(1,1,1)$ and
$v\,(1,\eta,\eta^\ast)$.\label{TabS4full2}}
\end{center}
\end{table}

\begin{table}[H]
\begin{center}
\begin{tabular}{|c|ccc|cccc||cl|}
\hline
vev&& $Q_L$ & & $u_R$ & & $d_R$ & & $\begin{array}{c}\text{Number of}\\\text{PDMA}\end{array}$& $\begin{array}{l}\text{Mass}\\\text{spectrum}\end{array}$\\
\hline\hline
\multirow{18}{*}{\rotatebox{90}{$(1,i,0)$}}&& $\mathbf{3_1}$ & & $\mathbf{1_1}$ & & $\mathbf{1_1}$  & & $0$  & $(0,0,m_{u,d})$\\
\cline{5-10}
&&  & & $\mathbf{1_1}$ & & $\mathbf{2,\,1_1}$ & &  $0$ &  $\begin{array}{l}(0,0,m_u)\\(0,m_d,m_d^\prime)\end{array}$\\
\cline{5-10}
&&  & & $\mathbf{1_1}$  & & $\mathbf{3_i}$  & & $0$  &  $\begin{array}{l}(0,0,m_u)\\(0,m_d,m_d)\end{array}$ \\
\cline{5-10}
&&  & & $\mathbf{2,\,1_1}$  & & $\mathbf{1_1}$  & & $0$  & $\begin{array}{l}(0,m_u,m_u^\prime)\\(0,0,m_d)\end{array}$\\
\cline{5-10}
&&  & & $\mathbf{2,\,1_1}$  & & $\mathbf{2,\,1_1}$  & & $0$  & $(0,m_{u,d},m_{u,d}^\prime)$\\
\cline{5-10}
&&  & & $\mathbf{2,\,1_1}$  & & $\mathbf{3_i}$  & & $0$ & $\begin{array}{l}(0,m_u,m_u^\prime)\\(0,m_d,m_d)\end{array}$\\
\cline{5-10}
&&  & & $\mathbf{3_i}$  & & $\mathbf{1_1}$  & & $0$  & $\begin{array}{l}(0,m_u,m_u)\\(0,0,m_d)\end{array}$\\
\cline{5-10}
&&  & & $\mathbf{3_i}$  & & $\mathbf{2,\,1_1}$  & & $0$  & $\begin{array}{l}(0,m_u,m_u)\\(0,m_d,m_d^\prime)\end{array}$\\
\cline{5-10}
&&  & & $\mathbf{3_i}$  & & $\mathbf{3_j}$  & & $0$  & $(0,m_{u,d},m_{u,d})$\\
\cline{2-10}
&& $\mathbf{2,\,1_i}$  & & $\mathbf{3_i}$  & & $\mathbf{3_i}$ & & $1$ & $(0,m_{u,d},m_{u,d}^\prime)$\\
\cline{2-10}
&& $\mathbf{1_i}$  & & $\mathbf{3_i}$  & & $\mathbf{3_i}$ & & $2$ & $(0,0,m_{u,d})$\\
\hline
\end{tabular}
\caption{As in Table~\ref{TabS4full1}; for the vev  $v\,(1,i,0)$.\label{TabS4full3}}
\end{center}
\end{table}

\section{\label{sec:conclusions}Conclusions}

We have studied the possibility of generating the quark masses and CKM mixing in the context of three Higgs doublet models extended by a discrete $A_4$ or $S_4$ symmetry. Assuming that the Higgs fields are in the triplet (faithful) representation of the discrete group, we have shown that none of the possible vev alignments that corresponds to a global minimum of the scalar potential leads to phenomenologically viable mass matrices for the three generations of quarks of the Standard Model and, simultaneously, to a non-vanishing CKM phase. Clearly, these conclusions can be evaded by extending the field content with extra scalars and/or fermions.

Our analysis can be applied straightforwardly to the leptonic sector of the theory, if neutrinos are Dirac particles. In that case, one massless neutrino or lack of leptonic CP violation would not contradict current experiments.

\begin{acknowledgments}
The work of R.G.F. and J.P.S. was partially supported by Portuguese national funds through FCT - \textit{Funda\c{c}\~{a}o para a Ci\^{e}ncia e a Tecnologia}, under the projects PEst-OE/FIS/UI0777/2011 and  CERN/FP/116328/2010, and by the EU RTN project Marie Curie PITN-GA-2009-237920. The work of H.S. is funded by the European FEDER, Spanish MINECO, under the grant FPA2011-23596.
\end{acknowledgments}

\end{document}